\documentstyle[aps,prl,epsf]{revtex}

\input {epsf}

\begin{document}
%\preprint{draft}
\twocolumn[\hsize\textwidth\columnwidth\hsize\csname@twocolumnfalse\endcsname

\title{Hysteretic Depinning of Anisotropic CDW}

\author{Valerii M. Vinokur$^{(a)}$ and Thomas Nattermann$^{(b)}$}
\address{(a) Material
  Science Division, Argonne National Laboratory, Argonne, IL 60439\\
(b) Institut f\"ur Theoretische Physik, Universit\"at zu
  K\"oln, Z\"ulpicher Str. 77, D-50937 K\"oln, Germany}

\date{\today}

\maketitle

\begin{abstract}
We investigate the depinning transition in a dirty periodic medium
considering a model of layered charge density waves as a prototype
system.  We find that depinning from strong disorder occurs via a two stage
process, where,  first, the pinned system experiences a continuous
transition into
a plastically sliding state and  undergoes a second
sharp hysteretic transition into a coherently
moving 3D state at higher drives. In the weakly disordered system
the depinning into a coherently sliding state 
remains continuous.
\end{abstract}


\vskip1pc]
\narrowtext

Depinning of periodic structures such as charge density waves (CDW),
vortex and
domain wall lattices, and Wigner crystal
from a random pinning
potential under the influence of an external driving force is one of the
paradigms of condensed matter physics. All these systems share one
thing in common: many
of elastically coupled degrees of freedom
interacting with a quenched random environment. In this Letter we report
analytic results
on the nature
of the depinning transition in dirty periodic media using
CDW dynamics as a prototype model and discuss possible extensions to other
systems.

Two types of depinning  have been observed: a smooth
non-hysteretic transition with a unique pinning threshold, and %a
transport switching characterized by an abrupt
hysteretic transition into a sliding state \cite{CDWswitch},
\cite{kwok}. Smooth depinning described in terms
of critical behavior \cite{dfisher} follows from the
description of the above systems as classical field  associated with the
distortions
of the system.
Much of the switching behavior is explained by
the possibility of plastic deformations allowing the amplitude of CDW to
vanish along certain surfaces within the system\cite{str,copp}.
Recent experimental and numerical studies
\cite{bhatt,kwokpl,jensen,kv}
of vortex transport in HTS have demonstrated
that the depinning of the vortex lattice
can also be accompanied by plastic effects.

These latter findings suggest that plastic effects can play an essential
role in the depinning, and that the nonequilibrium steady state near
the transition resembles fluid-like motion.
On the contrary, at very high velocities well above the depinning transition
the
influence of disorder on the dynamics is suppressed and one can expect
coherent motion of an almost perfect solid periodic system. A problem of
separation
of these two different driven regimes has been addressed in \cite{kv} in
the context of vortex transport.
It was 
proposed that the driven
periodic medium subject
to sufficiently strong disorder undergoes a sharp
hysteretic {\it dynamic} transition from coherent motion with
almost perfect structure to fluid-like plastic dynamics upon
decreasing the drive at a second critical force $F_{f}$ force {\it well
above the pinning threshold $F_{T}$}. This was called {\it dynamic freezing}.
Upon increasing the
driving force from the pinned state the vortex lattice starts to
slide at $F=F_{T}$, this
depinning being followed by the multiple plastic effects, and
elastic motion  recovers at $F=F_f>F_{T}$.
This concept received strong support both from earlier observations
of plastic effects \cite{bhatt} and from the subsequent transport measurements
on MoGe
superconducting films\cite{kapitulnik}.

The prediction of the possibility of a dynamical
phase transition in the driven state \cite{kv} was later expanded by nice
scaling
arguments onto three dimensional CDWs\cite{bf}.  The
properties of the driven coherent phase were examined and
discussed in \cite{gl,zimanyi,ryu,cbfm,brmcomment}.
Yet a number of unresolved problems remains. The fundamental issue is
the nature of the depinning transition (continuous vs. switching), and the
question under what conditions either type occurs.  How would
dynamic freezing evolve with decreasing strength of the disorder?
What are the conditions for the existence of the plastic flow regime as
the state intermediate between the pinned and coherently moving states?  Is
it possible to have depinning directly to a coherently moving state?

In this Letter we address these questions using CDW
transport as an example system. We consider a model of an anisotropic
CDW\cite{natthk}
%introduced by one of the coauthors
and develop a nonperturbative self-consistent
description of the dynamic transitions in driven dirty systems.
We find that if disorder is sufficiently strong, depinning occurs
in two stages: first the CDW depins in a driven decoupled state where 2D
CDWs in each layer slide independently, and second, upon further
increasing the
driving force
the system experiences a second transition into a coupled coherently
moving 3D phase.
%Dynamic coupling generally exhibits hysteretic switching behavior.
This "sequential depinning," corresponds to a
dynamic freezing transition scenario proposed in \cite{kv}, where the
periodic system first depins into a plastically moving state and then upon
further increase of the drive experiences a transition into a coherently
moving dynamic state. In the system with weak disorder decoupling
does not occur.

The paper is organized as follows. First we describe the model and derive
a selfconsistent equation for the shear modulus. Then we analyze the cases of
weak and strong pinning, and determine the disorder induced dynamic
decoupling (the instability point) as the current at which the
onset/disappearance of the coupling occurs.  In conclusion we construct the
general dynamic phase diagram for periodic structures driven through
quenched disorder.

%{\em Description of the  model}

The overdamped
dynamics of an anisotropic layered  CDW
%\cite{natthk}the system
are governed by the equation (CDW moves
along the $x$-direction):
           \begin{eqnarray}\nonumber
             \lambda{\dot{\phi}}_i({\bf x},t)
              =\gamma\nabla^2 \phi_i({\bf x},t)+
                \gamma\mu_0\left[\sin(\phi_{i+1}-\phi_i)+\right.\\
               \left.\sin(\phi_{i-1}-\phi_i)\right]+
               F+\gamma V \sin(\phi_i-\alpha_i({\bf x})).
\label{eq:discrmotion}
             \end{eqnarray}
Here $\phi$ is a CDW phase (displacement field), $\lambda$ is a friction
coefficient, $v$ is the average velocity of CDW, $\gamma$ is the elastic
constant, $\mu_0$ is the anisotropy parameter characterizing layer coupling,
$V$ is the strength of the random potential, $\alpha({\bf x})$ is a random
phase, $i$ is a layer index, and ${\bf x}$ is a $D$-dimensional vector in the
layer.  In the CDW models $\alpha$ comes from the backscattering part of 
the impurity potential: $\alpha_i({\bf x})=2{\bf k}_F({\bf x}+{\bf z_0}ia)$,
where ${\bf k}_F$ is the Fermi vector, and ${\bf r}=({\bf x},{\bf z_0}ia)$ 
is the coordinate of the impurity (${\bf z_0}$ is the unit vector in the $z$ 
direction).  Since the in-plane pinning correlation length exceeds the 
impurity spacing the random phase $\alpha$ is commonly considered as a random 
variable homogeneously distributed in the interval $[0,2\pi]$ 
\cite{CDWswitch}.

If the anisotropy parameter $\mu_0$ is large one recovers the continuous
limit,
then $i\to z$ and the term in square parenthesis becomes simply
$\partial^2\phi
/\partial z^2$.
The  related Hamiltonian %of the system
has the form:
     \begin{equation}
      %\nonumber
       {\cal H}=\gamma\!\int\!d^Dx\frac{dz}{a}
       {\left[\frac{1}{2}\left(\nabla\phi\right)^2\!+
           %\right. \\
           %\left.
       \frac{1}{2}\mu_0(\partial_z\phi)^2\!+
        V({\bf x},z,\phi)\right]} \label{eq:hamiltonian}
       \end{equation}
where we set $a=1$.
%As is well known the resulting continuous
This $D+1$-dimensional system experiences a continuous depinning transition at
critical force
$F_{c,d}=\gamma\left({V^4/\mu_0}\right)^{1/(4-d)}$,
$\mu_0\gg 1$, and the pinning correlation length is
$\xi_{c,d}=\left({V^4}/{\mu_0}\right)^{-{1}/{2(4-d)}}%\label{eq:xi_c}.
$, $d=D+1$.
If  $\mu_0\to 0$ the decoupling transition occurs in a static 3D system at
$\mu_0<\mu_{min}=V^2$.
Note that in a decoupled limit the equation of motion (\ref{eq:discrmotion})
becomes isotropic and $\mu_0\rightarrow 1$, $d\rightarrow D$.

To obtain a coarse-grained description in terms of a slowly varying
$\phi^<$ part of the phase
%. To this end we have to
we integrate out its fast
component $\overline{\phi}=\frac{1}{t_0}\int_0^{t_0}\phi(t)dt$, $t_0=l/v$, $l$
is
the CDW period
(hereafter we will drop the bar).  %
%Then one can write
%$\sin(\phi_{i+1}-\phi_i)\equiv\sin\phi=\sin(\phi^<+\phi^>)=\sin\phi^<\cos\phi^>
%+\sin\phi^>\cos\phi^<$
%(note that $\phi^>\ll 1$).  Carrying out the coarse-graining operation one
%finds
%$\overline{\sin\phi^>}=0$ and $\cos\phi^>\approx 1$.
The coarse graining procedure is straightforward in case of weak (as
compared to coupling) disorder, $\mu_0\gg \sqrt{V}$, where phase variations
from layer to layer are small and $\sin (\phi_{i+1}-
\phi_i)\approx (\phi_{i+1}-\phi_i)$.
Going over to a continuous description
one arrives at
the coarse-grained equation of motion for the slowly varying part of the
phase:
           \begin{equation}
              \lambda\dot{\phi}({\bf
                x},z)=\gamma\left(\nabla^2+\mu\partial_z^2\right)\phi-\lambda
              v\partial_x\phi+F_p({\bf x},z),\label{eq:motion}
            \end{equation}
where $\mu\approx\mu_0$ and $F_p$ is the effective random force (random
mobility-like term)
originating from the  random $\sin$ field
(analogous to \cite{stefan})
with the correlator defined as
$\langle F_p({\bf x},z)F_p({\bf
x}^{\prime},z^{\prime})\rangle=\Delta(v)\delta({\bf x}-{\bf
x}^{\prime})\delta(z-z^{\prime})$
with \cite{cbfm}, \cite{brmcomment}
          \begin{equation}
            \Delta(v)=\left\{
               {  \frac{(V\gamma)^4}{(\lambda v)^2} \;,
             \;\;\; v\gg\gamma\sqrt{\mu}/\lambda \atop
               \frac{V^4(\lambda v)^2}{\mu^2}\;, \;
             v\ll\gamma\sqrt{\mu}/\lambda }
             \right.
              \label{eq:delta}
           \end{equation}

%%%%%
We omitted the coarse-grained pinning potential term $f_p(\phi)$
and the disorder-induced KPZ term: it can be shown
\cite{stefan} that in the perturbative high velocity limit those terms
are irrelevant.

We are interested however in the strong disorder regime where the static CDW
is decoupled.  In this case the coarse-graining procedure is more subtle.
To carry it out note that in the high--velocity
case, $v\gg\gamma\sqrt{\mu}/\lambda\,,\; \Delta (v)$ does not depend
on the elastic constants $\gamma$ and the anisotropy parameter $\mu$
(provided the strength of disorder $\gamma V$ is fixed).
%This is due to the fact, that the response functions enter $\Delta (v)$
%only by its dissipative term $1/i\lambda v$.
This reflects the fact that at high velocities $\Delta (v)$ is controlled
by the dissipative part $1/i\lambda v$ of the response function
and enables us to carry out
%Consequently, the derivation of 
the coarse graining procedure also in 
%description in
the case of weak
anisotropy $\mu_0\ll\sqrt{V}$
(i.e.when a replacement of $\sin (\phi_{i+1}-\phi_i )$ by
$\phi_{i+1}-\phi_i$ is not possible).  The coarse-grained equation of motion
reads ($v\gg\gamma\sqrt{\mu}/\lambda$):
%{\it why $v \gg \Delta (v)$?, in the next line a dot was missing}

%the high velocity case can be repeated also
 %resulting in the following equation of motion
         \begin{eqnarray}
            \lambda\dot{\phi}_i({\bf x}) & = &
            \gamma {\bf \nabla}^2\phi_i({\bf x}) +
            \gamma\mu_0\left[ \sin (\phi_{i+1}-\phi_i )\right.
            \nonumber\\
            & + &  \left. \sin (\phi_{i-1}-\phi_i)\right]
            -  \lambda v\partial_x\phi_i +
            F_{p,i}({\bf x}).\hskip1.3cm (3')\nonumber
         \end{eqnarray}
It is convenient to rewrite the Eq. (3$^{\prime}$) in the form
         \begin{eqnarray}
            \lefteqn{\sum\limits_jG_{ij}^{-1}\phi_j = \gamma\mu_0\left[
            \sin{(\phi_{i+1}-\phi_i )}+\sin{(\phi_{i-1}-\phi_i)}\right]}
            \hspace{1cm}\nonumber\\
            & & - \gamma\mu\left[ \phi_{i+1}+\phi_{i-1}-2\phi_i\right]
            +F_{p,i}({\bf x})+\varepsilon_i({\bf x})
            \label{eq:solution3'}
         \end{eqnarray}
with
         \begin{eqnarray*}
             \lefteqn{G_{ij}^{-1}(t)= (\lambda\partial_t+\lambda v\partial_x-
             \gamma{\bf \nabla}^2+2\gamma\mu )\delta_{ij}}\hspace{4.5cm}\\
             & & - \gamma\mu(\delta_{i,i+1}+\delta_{i,i-1}), 
         \end{eqnarray*}
where $\varepsilon_i({\bf x})$ denotes a source term, which wil be sent
to zero at the
end of the calculations. One can now solve eq. (\ref{eq:solution3'})
iteratively, generating an infinite number of tree diagrams with either 
$F_{p,i}({\bf x})$ or $\varepsilon_i({\bf x})$ at the ends of each 
branch, and averaging subsequently
%the expression 
over the random force keeping only the linear
in $\varepsilon_i({\bf x})$ terms. To find the self-consistent equation for 
$\mu$ we note that the self-consistency condition requires that the 
phase-containing terms in
the r.h.s. of eq. (\ref{eq:solution3'})
canceled each other. After that we obtain in the lowest order in $\mu_0$:
        \begin{equation}
           \mu=\mu_0\exp\left[-\frac{1}{2}\langle\left(\phi({\bf x},z,t)-
           \phi({\bf x},z+a,t)\right)^2\rangle\right].\label{eq:mu}
       \end{equation}
Note also that since the correlations of $F_{p,i}({\bf x})$ are Gaussian the  
higher order cumulants do not appear within this scheme.

The correlation function in the exponent is calculated with the Hamiltonian
from
(\ref{eq:hamiltonian})
with $a=1$:
%%%%%%%%

        \begin{eqnarray}
           \nonumber
            C(\tilde{v},\mu,\Delta(v))=\langle\left(\phi({\bf x},z,t)-
             \phi({\bf x},z+a,t)\right)^2\rangle\\
               =\frac{1}{\gamma^2}\!\int\!
               \frac{d^3{\bf k}}{(2\pi)^3}\frac{\Delta(v)(1-\cos
               k_za)}{\left(k_x^2+k_y^2+\mu
              k^2_z\right)^2+\lambda^2v^2k_x^2/\gamma^2}
             \label{eq:corr}
           \end{eqnarray}
The main contribution comes from the maximal $k_z$, therefore we can replace
$k_z\rightarrow\pi/a$ in the integrand. Then in the 3D case one arrives at
        \begin{equation}
          C(\tilde{v},\mu,\Delta(v))=\frac{\Delta(v)}{(4\pi)^3
           \gamma^2\mu^{3/2}}
            \left[\left(\tilde{v}^2+4\pi^2\mu\right)^{1/2}-\tilde{v}\right],
          \label{eq:corr(v)}
        \end{equation}
where $\tilde{v}=\lambda v/\gamma$.  Outside the critical region
$\tilde{v}\approx F/\gamma$, and within the critical region
${\tilde v}\simeq(F-F_c)^{1-(4-d)/6}/\gamma$\cite{narfish}.
Note that in the limit of very large velocities Eq.(\ref{eq:corr})
provides
the expected behavior $\mu\rightarrow\mu_0$, and the system always
remains coupled.

Now comes the central point of our discussion: solving the self-consistent
equation for $\mu$.  The disappearance of the solution to Eq.
(\ref{eq:mu}) implies
decoupling of the system.  To capture the transition we will be
seeking for the moment of the first disappearance of the solution.
For the sake of simplicity we can replace
the expression in square brackets in (7) by
${2\pi^2\mu}/{\sqrt{\tilde{v}^2+4\pi^2\mu}}$, which gives the same
asymptotics in the limit of small and large velocities.
As we will shortly see in the case of strong
disorder the
decoupling transition occurs at large velocities, whereas in the weak
disorder
case there is no decoupling.
%Respectively,
%we will deal separately with the large and small velocity limits verifying
%afterwards the consisteency of the coresponding assumptons.

{\em
Strong disorder: $V^2>\mu_{\circ}$}.
In the limit of large velocities  Eq.(\ref{eq:mu}) assumes the form:
     \begin{equation}
       \mu=\mu_{\circ}\exp\left(-\frac{V^4}{32\pi\tilde{v}^3\mu^{1/2}}\right).
       \label{eq:mularge}
      \end{equation}

To find the point $\mu_c$ of the disappearance of the solution we derivate
both sides of (\ref{eq:mularge}), obtaining the condition
%\begin{equation}
$V^4=64\pi\tilde{v}^{3}\mu^{1/2}$
%\label{eq:largederivative}
%\end{equation}
and find
    \begin{equation}
     \mu_c=\mu_{\circ}/e^2 ,\,\     
\tilde{v}_c=\left(\frac{e}{64\pi}\frac{V^4}{\sqrt{\mu_{\circ}}}\right)^{1/3}
      \label{eq:mucritlarge}
     \end{equation}

The last task to complete our calculation is to verify that the critical
decoupling velocity indeed falls into a large velocities
interval.  To this end note that the large velocity condition  $\tilde{v}_c\gg
\sqrt{\mu}$ reduces at the instability point to
$V^2\gg8\sqrt{\pi}\mu_{\circ}/e^2$ which is just the condition of the
strong disorder assumed and therefore our assumption is justified (one has to
bear in mind that in our dimensionless units $\mu\ll V^{2}\ll 1$).
The critical velocity
(\ref{eq:mucritlarge}) agrees with the result of \cite{bf} suggested
by nice scaling arguments if one substitutes $\Delta(v)$ from
(\ref{eq:delta}) instead of the unspecified mean squared
pinning strength $g$ of\cite{bf}.

To understand the meaning of the obtained result let us construct the
$F-v$ dependence starting with the ascending branch.  Below the 2D critical
force $F_c^{2D}$
the {\it decoupled} system remains pinned.  At $F=F_c^{2D}$ the system
undergoes a smooth depinning transition into the plastically moving state
in which the system remains decoupled and each
layer moves independently.  Upon further increase of the drive, the mean
velocity
of the CDW reaches the critical value and the system gets coupled into a 3D
moving state.  Note that since in the 3D regime pinning force experienced 
by the moving CDW, $F_p^{3D}(v)$, is less then the corresponding pinning force
in the 2D regime: $F_p^{3D}(v)\simeq
F_c^{2D}(v)/\sqrt{L_{\perp}(v)}< F_p^{2D}(v)$, where $L_{\perp}(v)$ is 
the (velocity dependent) correlation
length across the layers, the pinning correction to the 3D velocity is {\it
smaller} than the corresponding correction to the 2D velocity.
%depinning force $F_c^{3D}\simeq
%F_c^{2D}/\sqrt{L_{\perp}}< F_c^{2D}$, where $L_{\perp}$ is the correlation
%length across the layers, the pinning correction to the 3D velocity is {\it
%smaller} than the corresponding correction to the 2D velocity.  
As a result the 3D branch of the $F-v$ dependence
lies {\it above} the $v_{2D}(F)$ curve and the transition from the plastic
to the coupled elastic motion at $v=v_c$ upon increasing drive acquires
an {\it abrupt switching character} (see Fig.1).  Going down from the high
velocities the system follows first the elastic 3D behavior with
$\mu \approx \mu_0$, and then as velocity decreases to $v=v_c$, the system
decouples and jumps down to the 2D branch corresponding to the plastic
motion (see Fig. 1a) at $F=F_{c(down)}<F_{c(up)}$.  Therefore
in the limit of strong pinning the transition from plastic to elastic
motion is a switching hysteretic transition.

An immediate reservation regarding the proposed %above 
scenario is in order.  The
above picture suggests that the real {\it unique} $F-v$ characteristic
describing 
the dynamic behavior of the system is the  
$S$-shape-like curve, and, accordingly, up- and down-switchings
occur at the instability points where $|dv/dF|= \infty$.  To derive 
rigorously this kind of the $v(F)$ dependence a careful analysis in the 
vicinity of
the critical point $v_c$ accounting for the interaction between the
2D and 3D modes and nonlinear pinning effects is needed.  Leaving the
detailed derivation for a forthcoming publication we point out that the
%another
related scenario where the critical points merge and $v(F)$
retains only the inflection point near $v=v_c$ is also possible.  
In this degenerate case
the plastic--elastic dynamic transition becomes non-hysteretic and can be
detected by the position of the inflection point in $v(F)$ dependence.

Turning now to the case of {\it weak disorder, $V<\mu^2$,}one can be 
easily verify, that in this case
there will be always a non--zero solution for $\mu$ (note
that for weak disorder $F_c^{3D}=V^4/\mu <F_c^{2D}=V^2$).  Combining 
this observation with the Eq. (\ref{eq:mucritlarge}) we conjecture the 
$v-V$ diagram shown in Fig 2.

Our discussion was restricted to $T=0$, but the form of the  F-v curve 
(S-shape, for example) is determinbed by the intrinsic dynamic 
properties of the system and is to be stable 
with respect to thermal effects (as long as the the latter leave the periodic
structure intact). Another point is that although our
consideration was focused on an anisotropic CDW model, the form of 
the coarse-grained random force we used is not specific to CDWs but 
seem to be generic for any periodic structure subject to quenched 
disorder \cite{brmcomment,stefan}. We 
expect therefore that the above ideas apply to a general case of  
periodic media driven
through quenched disorder at finite temperatures.  In particular, one 
can view the 
``strong depinning'' switching scenario as the ``zero-temperature 
projection'' of the sequential depinning or non-equilibrium
freezing transition in the vortex system subject to strong
disorder proposed in \cite{kv}. The instability force transcribes
into a freezing force introduced for a moving vortex lattice and the
independent layered motion of 2D CDWs maps onto a
regime of plastic flow at the intermediate currents $j_{c}<j<j_{f}$ 
(see inset in the Fig 1a). 
The details of the finite temperature behavior as
well as the quantitative transcription of the ideas developed for the
anisotropic CDW onto a general case of periodic driven media will be
presented elsewhere.  

In conclusion, we described the depinning behavior in the driven
periodic structures using the model of the layered CDW as a prototype
system.  We have found that in case of strong disorder where the
static state is decoupled the depinning occurs via a two stage
process.  First the pinned system experiences a continuous depinning into
a plastically sliding state and at higher drives undergoes a second
sharp hysteretic transition into a coherently
moving 3D state.  The second transition is identified with the freezing
transition proposed in \cite{kv}.  For weak disorder depinning occurs 
into a coupled state.

It is a great pleasure to thank I. Aranson and S. Scheidl for many useful
discussions and G. Crabtree for critical reading of the manuscript.
This work was supported from Argonne National
Laboratory through the U.S.  Department of Energy, BES-Material
Sciences, under contract No. W-31-109-ENG-38 and by the NSF-Office of
Science and Technology Centers under contract No.  DMR91-20000 Science
and Technology Center for Superconductivity.

\widetext
\begin{figure}
\epsfxsize=3.4 in
\epsfbox{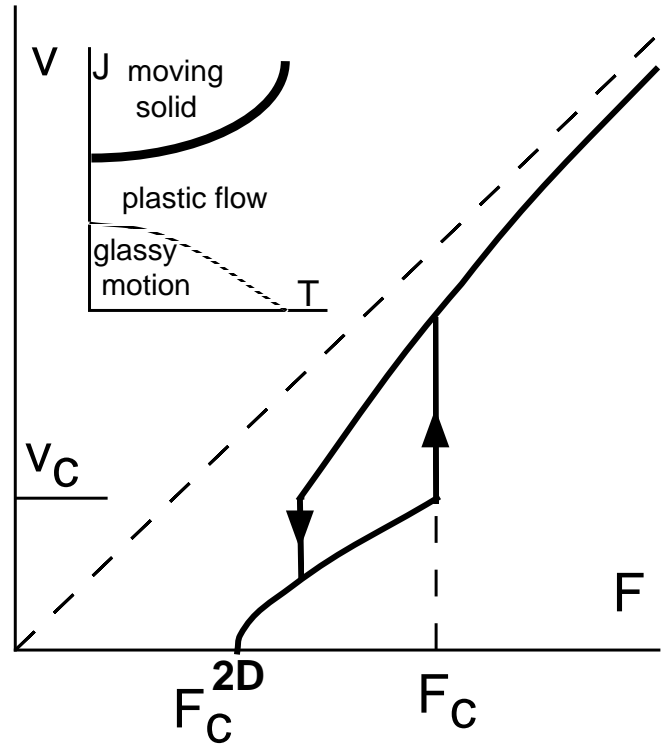}
%\epsfbox{fig11.ps}
%\epsffile{Fig11.ps}
\narrowtext
\caption{
$v-F$ transport characteristic for the strong pinning at $T=0$.  The dashed
line displays viscous behavior in the absence of pinning.  {\it
Inset:} Dynamic phase diagram for periodic medium driven through
strong disorder. The dashed line denotes the critical depinning force
$j_{c}(T)$
at which crossover from creep plastic flow takes place.  The solid line
denotes the switching-like freezing transition at $j_{f}(T)$.
}

\label{fig.down}
\end{figure}

\begin{figure}
\epsfxsize=3.4 in
\epsfbox{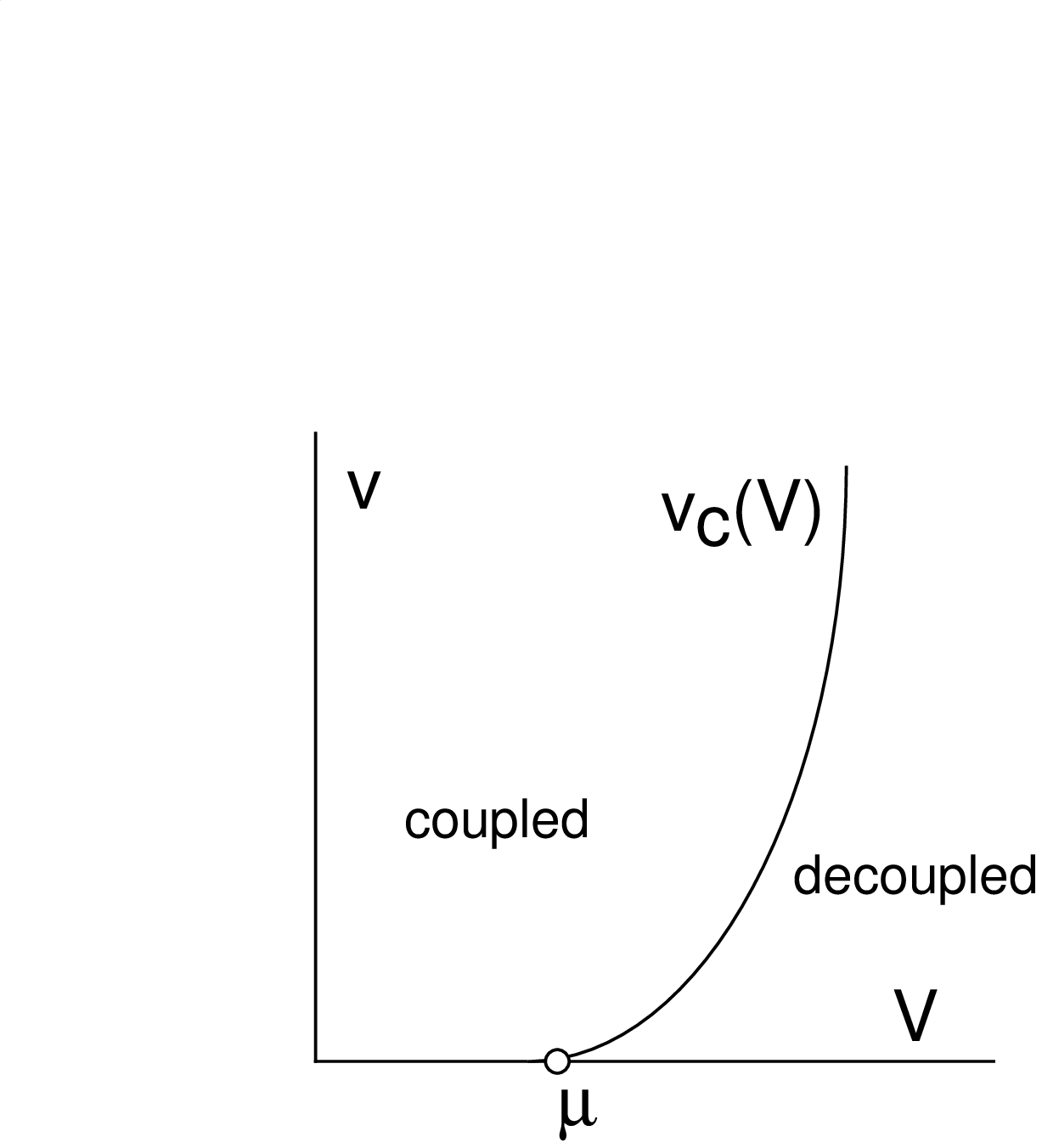}
%\epsffile{Fig2.ps}
\narrowtext
\caption{
$v-V$ phase diagram: the system recovers coupling at $v>v_{c}(V)$, 
the large velocity dependence of $v_{c}(V)$ is given by
Eq.(\ref{eq:mucritlarge})o}

\label{fig.phase}
\end{figure}
\end{document}